\documentclass{article}

\usepackage{arxiv}

\usepackage[T1]{fontenc}    
\usepackage{url}            
\usepackage[ toc, page, title, titletoc ]{appendix}
\usepackage[ruled,noline]{algorithm2e}
\usepackage{amsmath}
\usepackage{amsfonts}
\usepackage{booktabs}
\usepackage[colorinlistoftodos]{todonotes}
\usepackage{hyperref}
\usepackage{lineno}
\usepackage{mathrsfs}
\usepackage{float}
\usepackage{url}            

\usepackage{etoolbox}
\providetoggle{fullBackground}

\usepackage{parskip}  

\DeclareFontFamily{U}{wncy}{}
\DeclareFontShape{U}{wncy}{m}{n}{<->wncyr10}{}
\DeclareSymbolFont{mcy}{U}{wncy}{m}{n}
\DeclareMathSymbol{\comb}{\mathord}{mcy}{"58} 

\usepackage{accents}
\newlength{\dhatheight}

\title{Reducing the Sampling Burden of Fourier Sensing
with a Non-rectangular Field-of-View}

\author{
  Nicholas Dwork\thanks{www.nicholasdwork.com, nicholas.dwork@cuanschutz.edu} \\
  Department of Biomedical Informatics \\
  University of Colorado | Anschutz Medical Center
  \And
  Erin K. Englund \\
  Department of Radiology \\
  University of Colorado | Anschutz Medical Campus
  \And
  Alex J. Barker \\
  Department of Radiology \\
  University of Colorado | Anschutz Medical Campus
}

\begin{document}
\maketitle

\begin{abstract}
With Fourier sensing, it is commonly the case that the field-of-view (FOV), the area of space to be imaged, is known prior to reconstruction.  To date, reconstruction algorithms have focused on FOVs with simple geometries: a rectangle or a hexagon.  This yields sampling patterns that are more burdensome than necessary.  Due to the reduced area of imaging possible with an arbitrary (e.g., non-rectangular) FOV, the number of samples required for a high-quality images is reduced.  However, when an arbitrary FOV has been considered, the reconstruction algorithm is computationally expensive.  In this manuscript, we present a method to reduce the sampling pattern for an arbitrary FOV with an accompanying direct (non-iterative) reconstruction algorithm.  We also present a method to decrease the computational cost of the (iterative) model-based reconstruction (MBR) algorithm.  We present results using MRI data of an ankle, a pineapple, and a brain.
\end{abstract}

\keywords{Fourier Sensing \and Magnetic Resonance Imaging \and Reconstruction \and Sampling Theory}

\section{Introduction}

Fourier sensing imaging devices collect data in the frequency domain; some form of inverse Fourier transform is then required to reconstruct the image.  Examples of Fourier sensing systems include Magnetic Resonance Imaging , Computed Tomography, Optical Coherence Tomography, Synthetic Aperture Radar, and Radio Astronomy.  For these devices, it is commonly the case that the field-of-view (FOV), the area of space to be imaged, is rectangular.  Assuming the sampling pattern is a Cartesian grid, this rectangular FOV dictates the spacing between samples required to satisfy the Nyquist-Shannon sampling theorem: the spacing between grid points in the frequency domain is the inverse of size of the FOV in the corresponding dimension \cite{nishimura1996principles}.  This sampling pattern is commonly referred to as \textit{fully-sampled} \cite{sumpf2011model, kajbaf2013compressed, baron2018rapid, cole2021fast}.

\begin{figure}[ht]
  \centering{}
  \includegraphics[width=0.8\linewidth]{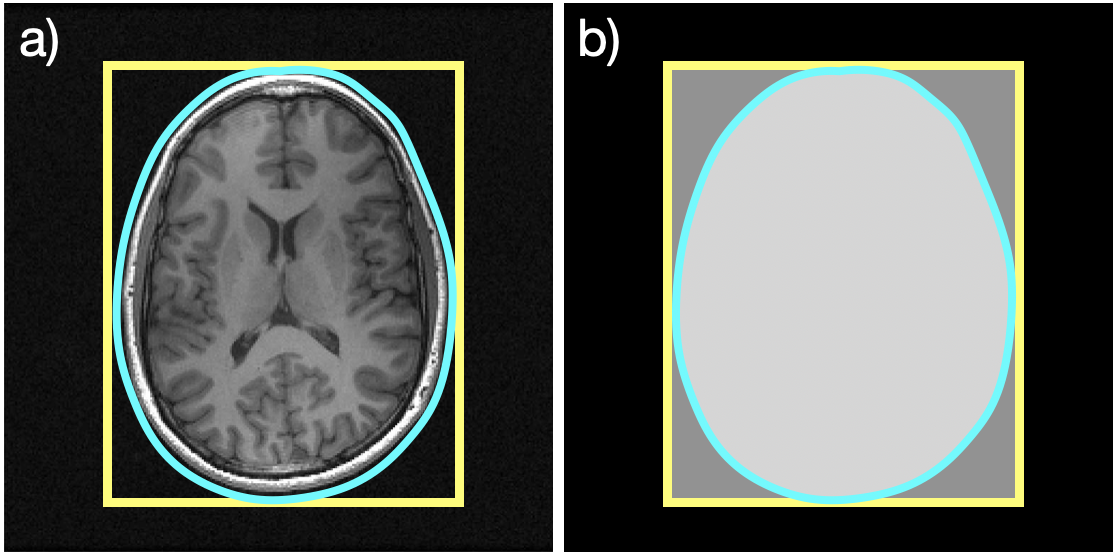}
  \caption{ An axial slice of a brain: a) the yellow rectangle and the cyan contour represent a rectangular and non-rectangular FOV, respectively, and b) a depiction of the reduced area in the non-rectangular FOV. }
  \label{fig:brainAreaFOV}
\end{figure}

It is often the case, though, that the support of the object (i.e., the area where the object is present) to be imaged is significantly smaller than the FOV.  Consider the axial slice of a brain shown in Fig. \ref{fig:brainAreaFOV}a; the yellow rectangle and cyan contour represent the boundaries of a rectangular and non-rectangular FOV, respectively.  As shown in Fig. \ref{fig:brainAreaFOV}b, the area enclosed by the cyan contour is smaller than that of the yellow rectangle.  It would seem, then, that fewer samples should be required for a high-quality image of the reduced FOV.

Past works have attempted to develop reconstruction algorithms that are able to generate high-quality images from fewer samples with a reduced non-rectangular FOV.  For a hexagonal FOV, one can sample with a hexagonal sampling pattern and use the inverse hexagonal FFT (HexFFT) to reconstruct the image \cite{dudgeon1983multidimensional, bracewell1995two}.  The HexFFT can be implemented with standard and computationally efficient FFT implementations \cite{birdsong2016hexagonal}.

Though useful, the HexFFT is restricted to hexagonal FOVs.  We would like to consider a more general FOV that need not be a hexagon (or even a polygon).  In \cite{samsonov2004pocsense}, Samsanov et al. describe POCSENSE, a method for taking advantage of an arbitrary FOV.  The method is iterative, which is computationally expensive, and reconstructs the entire rectangular FOV under the constraint that the pixels outside of the support be $0$.

In this manuscript, we present a direct (non-iterative) method for reconstructing an image from a reduced sampling pattern generated from a non-rectangular FOV\footnote{Note that an early version of this work was presented at the 2024 IEEE conference on Computational Imaging and Synthetic Apertures.}.  We will present examples where images of high-quality are reconstructed from a reduced sampling pattern using Magnetic Resonance Imaging (MRI) data of an ankle, a pineapple, and a brain.  We will extend this technique to reconstruct high-quality images from parallel MRI data, where multiple coils (i.e., antennas) simultaneously image the object.  Finally, we will present several avenues for future work.

\section{Background}

For the purposes of discussion, we assume that the non-rectangular FOV is closed and connected, and that it is known \textit{a priori}.  With MRI, the FOV is commonly determined from a \textit{localizer}, a low-quality image generated from data collected with a fast scan used for the purposes of identifying the relevant anatomy and selecting the FOV.  With radio astronomy, it may be known that the object of interest only encompasses a small volume of space.

The two-dimensional (2D) Fourier transform of a function $I:\mathbb{R}^2\rightarrow\mathbb{C}$ is
\begin{equation}
  \mathcal{F}\{I\}\left(k_u,k_v\right) = \iint_{\infty}^{\infty} I(u,v) e^{ -i 2\pi k_u u + k_v v} \, du\, dv.
\end{equation}
With Fourier sensing devices, typically values of $\mathcal{F}\{I\}$ are measured at individual spatial frequency coordinates $(k_u,k_v)$.  With MRI, values of $\mathcal{F}\{\sigma I\}$ are measured, where $\sigma:\mathbb{R}^2\rightarrow\mathbb{C}$ quantifies the sensitivity of the antenna as a function of space.  Parallel MRI simultaneously collects data with multiple antennas.

\begin{figure}[ht]
  \centering{}
  \includegraphics[width=0.8\linewidth]{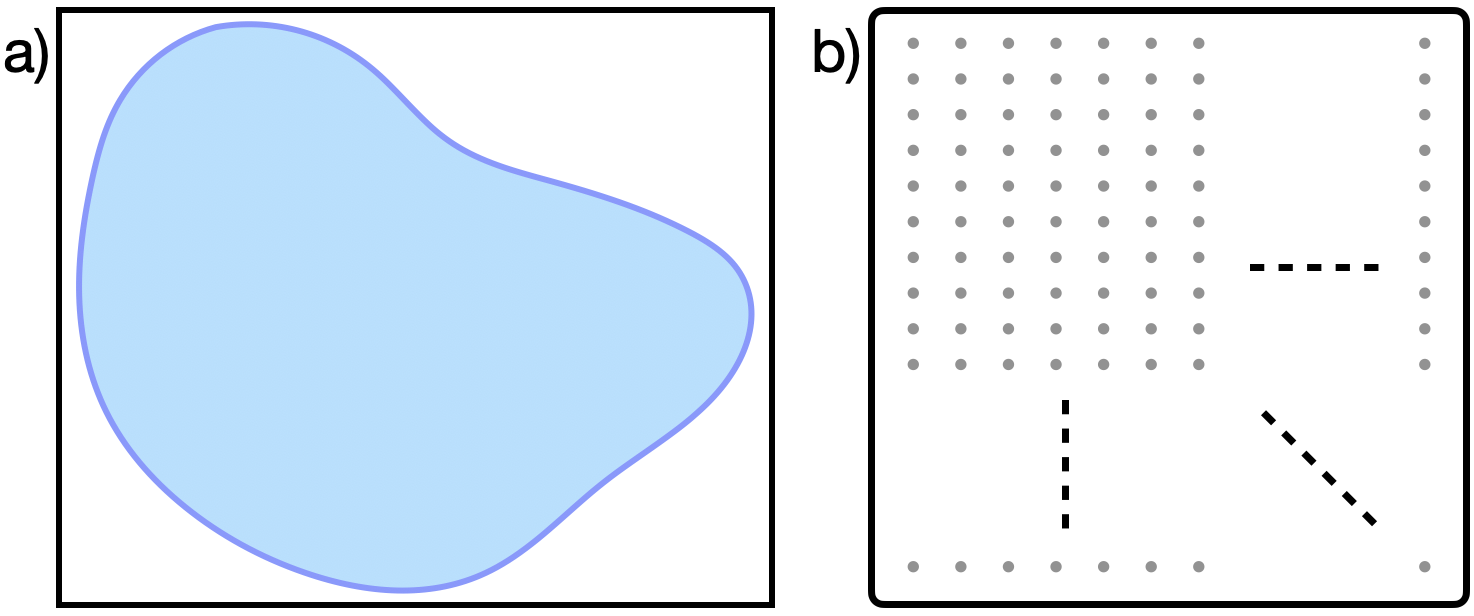}
  \caption{ A depiction of a non-rectangular FOV and its sampling pattern.  The dark blue contour in (a) shows the boundary of the non-rectangular FOV, and (b) shows the full sampling pattern associated with a rectangular FOV. }
  \label{fig:fullSamplingPattern}
\end{figure}

To satisfy the Nyquist-Shannon sampling theorem with a FOV of dimensions $\text{FOV}_u\times\text{FOV}_v$, the sampling pattern is a Cartesian grid centered on the $(0,0)$ frequency and the spacing between samples should be greater than $1/\text{FOV}_u$ and $1/\text{FOV}_v$ in the $u$ and $v$ dimensions, respectively.  The resolution of the resulting image is determined by the number of samples.  Such a sampling pattern for the arbitrary object of Fig. \ref{fig:fullSamplingPattern}a is depicted in Fig. \ref{fig:fullSamplingPattern}b.

The insight that leads to the method of this manuscript is that any non-rectangular image can be decomposed into the sum of two images with smaller supports.  For example, the object of Fig. \ref{fig:fullSamplingPattern}a can be decomposed as shown in Fig. \ref{fig:imgParts}b-c.  Remarkably, as long as we are careful about how we separate the image into its parts, the sampling pattern required to satisfy the Nyquist theorem for the image's components has fewer samples than the fully-sampled pattern represented in Fig. \ref{fig:fullSamplingPattern}b.

\section{Methods}
\label{sec:methods}

In this section, we will first discuss how the reduced sampling pattern is generated.  We will then present a method for reconstructing the image from the reduced sampling pattern.

\subsection{Generating the Sampling Pattern}

\begin{figure}[ht]
  \centering{}
  \includegraphics[width=0.95\linewidth]{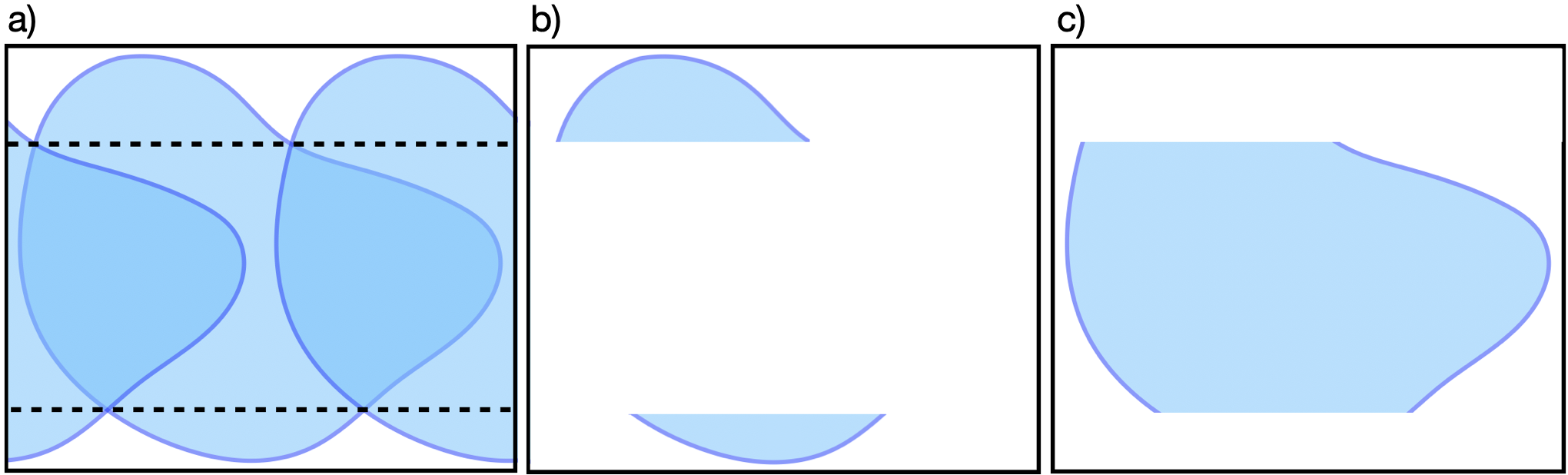}
  \caption{ A depiction of how a non-rectangular FOV is separated into components: a) shows an aliased version non-rectangular FOV of Fig. \ref{fig:fullSamplingPattern}a where the dashed lines separate the outer region of (b) and the inner region of (c). }
  \label{fig:imgParts}
\end{figure}

Consider the reconstruction created by applying the inverse DFT to a sampling pattern that consists only the even columns of the full sampling pattern (where the odd columns are filled with $0$ values).  The reconstructed image includes significant aliasing; it is the sum of the image with itself circularly shifted by half the rectangular FOV, as shown in Fig. \ref{fig:imgParts}a.  For the example depicted, that the boundary of the original object intersects that of its aliased copy in four places.  The dashed horizontal lines cross these intersection points.  Let us denote the region consisting of that above the top dashed line as well as that below the bottom dashed line -- as the outer region (Fig. \ref{fig:imgParts}b), and the remaining areas -- the region within the two dashed lines -- as the inner region (Fig. \ref{fig:imgParts}c).  The data for the non-rectangular FOV sampling pattern was created by combining columns from a fully-sampled pattern and that with a FOV of that was reduced in the vertical (superior-inferior) direction; these data were collected with currently available clinical scanning protocols.  Any subject motion that occurs in between data collections leads to errors in the reconstruction.

The inner and outer regions can be identified as follows.  Let $S$ be a 2D array, the size of the desired image, that indicates the support based on the non-rectangular FOV.  That is, those elements of $S$ that correspond to pixels within the FOV have values of $1$, and all other elements have values of $0$.  The rows of $S$ in the inner region are identified by identifying those rows of $S + \tau_u(FOV/2)$ with any element equal to $2$; here, $\tau_u(FOV/2)$ represents a circular shift in the horizontal (or $u$) dimension.  The rows of $S$ in the outer region are the complement set of those in the inner region.  After the rows of the inner region are identified, the support of the inner region $S_{\text{inner}}$ is found by intersecting the inner region with the the non-rectangular FOV and its interior.  Performing the same procedure with the rows of the outer region yields the support of that region $S_{\text{outer}}$.

The aliased image of Fig. \ref{fig:imgParts}a can be reconstructed by performing an inverse DFT on using only the even columns of the full sampling pattern (Fig. \ref{fig:fullSamplingPattern}b).  Note that, in the aliased image (Fig. \ref{fig:imgParts}a), the outer region does not overlap itself.  Performing a Hadarmard product (i.e., a point-wise product, denoted by $\odot$) between the aliased image and $S_{\text{outer}}$ (Fig. \ref{fig:imgParts}a) yields an uncorrupted image of the outer region (Fig. \ref{fig:imgParts}b).  This shows that we can accurately reconstruct the outer region using every other column of the full sampling pattern.

Thus far, our sampling pattern consists of samples only from the even columns of the full sampling pattern.  We will now add additional samples into the sampling pattern to reconstruct the inner image depicted in Fig. \ref{fig:imgParts}c.  

\begin{figure}[ht]
  \centering{}
  \includegraphics[width=0.95\linewidth]{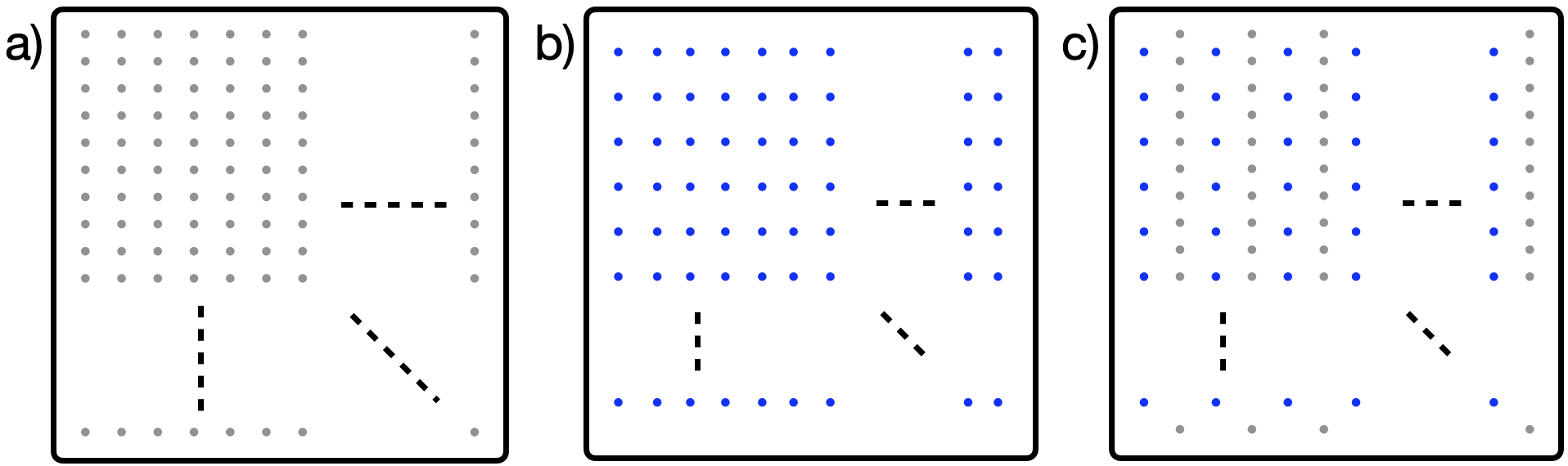}
  \caption{ Sampling patterns related to the FOV: a) the full sampling pattern based on the rectangular FOV, b) the full sampling pattern for the inner region, and c) the proposed sampling pattern for the non-rectangular FOV. }
  \label{fig:samplingPatterns}
\end{figure}

If we were going to create a full sampling pattern for the inner region, we would create a Cartesian grid with a horizontal spacing equal to that depicted in Fig. \ref{fig:fullSamplingPattern}b, since the horizonta; extent of the inner region is the same as that of the full non-rectangular FOV.  However, the vertical extent of the inner image is smaller than that of the non-rectangular FOV; thus, its samples would be separated further apart vertically.  The full sampling pattern would be a Cartesian grid with points separated horizontally by $1/\text{FOV}_u$ but with points separated vertically by an amount greater than $1/\text{FOV}_{v,\text{inner}}$; this is depicted in Fig. \ref{fig:samplingPatterns}b.  This Cartesian grid satisfies the Nyquist-Shannon sampling theorem.  Therefore, any pattern with a density of samples higher than that of this grid also satisfies the Nyquist-Shannon sampling theorem.  We replace all the odd columns of the full sampling pattern with the corresponding columns of the full sampling pattern for the inner region (Fig. \ref{fig:samplingPatterns}c); this still satisfies the requirements of the Nyquist-Shannon sampling theorem for the inner region.

Figures \ref{fig:samplingPatterns}c shows the proposed sampling pattern for the non-rectangular FOV; the gray dots are those that were used to reconstruct the outer image; the blue dots are the sample locations that will be added for the inner image.  The samples added into the pattern to reconstruct the inner image are positioned horizontally at the odd columns of the original fully-sampled pattern, but are separated by the inverse of the vertical extent of the inner region.  The increased vertical separation between samples is the reason why the new sampling pattern has a reduced number of samples when compared to the original full sampling pattern.

\subsection{Image Reconstruction}

In this subsection, we will first describe the direct reconstruction approach (a non-iterative approach for image reconstruction).  We will then discuss an iterative model-based reconstruction approach \cite{fessler2010model} where the reconstructed image is the solution of an optimization problem.

\subsubsection{Direct Reconstruction}

Let $I$, $I_{\text{outer}}$, and $I_{\text{inner}}$ denote the image, the outer region, and the inner region, respectively.  Let $F$ denote the Discrete Fourier Transform (DFT), let $\mathcal{F}$ denote the continuous Fourier transform, and let $F_{nu}^{(k)}$ denote the non-uniform DFT to the set of frequencies contained in the set $k$.  The $I_{\text{outer}}$ image is reconstructed from the even columns of the sampling pattern.  Then
\begin{align}
  \mathcal{F}(I) &= \mathcal{F}\left(I_{\text{outer}}\right) + 
    \mathcal{F}\left(I_{\text{inner}}\right) \nonumber \\
  \Rightarrow I_{\text{inner}} &= \mathcal{F}^{-1}\left\{ \mathcal{F}(I) - 
    \mathcal{F}\left(I_{\text{outer}}\right) \right\}.
    \label{eq:nonRectIFFT}
\end{align}
Equation \eqref{eq:nonRectIFFT} amounts to the reconstruction algorithm for $I_{\text{inner}}$, where the continuous Fourier transforms are replaced by the appropriate DFT \cite{greengard2004accelerating} (either gridding \cite{jackson1991selection, beatty2005rapid, dwork2023optimization} or inverse gridding \cite{pauly2005nonCartesian, rasche1999resampling}) .  Let $F_{nu}^{(k)}$ and $F^{\text{inv},(k)}_{nu}$ denote gridding and inverse gridding, respectively\footnote{Note that, in general, $F_{nu}$ is not invertible and $F^{\text{inv}}_{nu}$ is an approximation to its inverse.}.

Let $k_{\text{even}}$ and $k_{\text{odd}}$ denote the set of frequencies collected from the even and odd columns of the sampling pattern, respectively.
Let $k_{\text{all}}=k_{\text{odd}}\cup k_{\text{even}}$.  Let $k_{\text{inner}}$ denote the set of frequencies that would be required for a fully-sampled pattern of the inner region (Fig. \ref{fig:samplingPatterns}b).
Let $\mathcal{F}\{I\}_{\text{even}}$ and $\mathcal{F}\{I\}_{\text{all}}$ denote $\mathcal{F}\{I\}(k_{\text{even}})$ and $\mathcal{F}\{I\}(k_{\text{all}})$, respectively.
Let $S$ denote a 2D array of the size of the image with values of $1$ for those pixels within the non-rectangular FOV, and $0$ otherwise.  Let $S_{\text{outer}}$ and $S_{\text{inner}}$ denote 2D arrays indicating the support of the outer and inner images, respectively.
The direct (non-iterative) algorithm for reconstruction of an image with a non-rectangular FOV is presented in Alg. \ref{alg:nonRectRecon}.  This algorithm does the following: 1) reconstructs the outer region from the even columns, 2) interpolates all data onto the fully-sampled pattern for the inner region, 3) subtracts the Fourier values of the outer region from the samples of the inner grid, 4) reconstructs the inner region, and 5) sums the outer and inner regions together to create the final image.

\begin{algorithm}[ht]
    \protect\caption{Direct reconstruction with a non-rectangular FOV}
    \label{alg:nonRectRecon}

    \textbf{Inputs:}  $S$, $k_{\text{even}}$, $k_{\text{odd}}$, $\mathcal{F}\{I\}_{\text{even}}$, and $\mathcal{F}\{I\}_{\text{odd}}$

    $I_{\text{outer}} = S_{\text{outer}} \odot F^{-1}\left( \mathcal{F}\{I\}_{\text{even}} \right)$

    $F_{\text{inner,odd}} = \mathcal{F}\{I\}_{\text{odd}}$

    $F_{\text{inner,even}} = F^{(k_{\text{inner,even}})}_{nu} \left[ F^{\text{inv},(k_{\text{even}})}_{nu}\left( \mathcal{F}\{I\}_{\text{even}} \right) \right]$

    $F_{\text{inner}} = F_{\text{inner,odd}} \cup F_{\text{inner,even}}$

    $I_{\text{inner}} = F_{nu}^{(k_{\text{inner}})}\left[ F_{\text{inner}} - F^{\text{inv},(k_{\text{inner}})}_{nu}\left(I_{\text{outer}}\right) \right]$

    $I = S \odot \left( I_{\text{inner}} + I_{\text{outer}} \right)$

    \textbf{Output: } $I$
\end{algorithm}

With parallel MRI, where multiple coils simultaneously collect data at the same spatial frequencies, we can also use the direct algorithm.  To do so, we will determine the non-rectangular FOV for each sensitized image $\sigma^{(j)}\,I$; here, $\sigma^{(j)}$ denotes the sensitivity of the $j^{\text{th}}$ coil.  A simple approach would be to compute the union of these non-rectangular FOVs, and proceed to reconstruct each coil's image using Alg. \ref{alg:nonRectRecon}.  However, this would yield a sampling pattern that is likely to be more burdensome than is necessary.  Instead, we will individually identify each coil's non-rectangular FOV, determine the resulting sampling pattern for each individual non-rectangular FOV, and use the resulting sampling pattern with the most number of samples.

The image for each coil $I^{(j)}$ is reconstructed using Alg. \ref{alg:nonRectRecon}.  Then, each coil's image is multiplied by the support for that coil: $I^{(j)} := S^{(j)} \odot I^{(j)}$.  The images from all coils can then be combined into a single image using the method of Roemer et al. \cite{roemer1990nmr}.

\subsubsection{Model-based Reconstruction}

For typical Fourier sensing, where the data collected are the Fourier values of the image to be reconstructed, one can reconstruct the image by solving the following optimization problem: $\text{minimize} \| F_{nu}\,I - b \|$, where $\|\cdot\|$ represents a norm, $b$ represents the data collected, and $I$ is the optimization variable.  When Gaussian noise is present, which is what we will assume for the remainder of this manuscript, the $\ell_2$ norm provides an optimal reconstruction.  For the special case where the data collected is on a Cartesian grid centered on the $0$ frequency and satisfies the Nyquist-Shannon sampling theorem, $F_{nu} = F$ and the optimal reconstruction is $I^\star = F^{-1} b$.

When a non-rectangular FOV is provided with a corresponding support $S$, one can find the optimal estimate by solving
\begin{equation*}
    \text{minimize} \hspace{0.5em} \| F_{nu}\,I - b \|_2 \hspace{0.5em}
    \text{subject to} \hspace{0.5em} S^C \odot I = 0.
\end{equation*}
Here, $S^C = 1 - S$ is the complement of $S$.  For the case where the data is collected on a Cartesian grid, this problem becomes
\begin{equation*}
    \text{minimize} \hspace{0.5em} \| M_b \, F \, I - b \|_2 \hspace{0.5em}
    \text{subject to} \hspace{0.5em} S^C \odot I = 0,
\end{equation*}
where $M_b$ is a mask that indicates which data values of the Cartesian grid were collected.

In \cite{samsonov2004pocsense}, Samsonov et al. suggest using the Projection Onto Convex Sets (POCS) algorithm to solve this problem.  This approach iterates over the following two steps:  1) project onto the set of images that are consistent with the data collected, and 2) project onto the set of images where pixels outside of the FOV all have values equal to $0$.  While effective, this algorithm is computationally expensive because it is slow to converge.  It is also intuitively inefficient: we should not need to include pixels outside of the FOV in the optimization variable when we know their values are all $0$.  Instead, we propose solving the following optimization problem:  
\begin{equation}
  \text{minimize} \hspace{0.5em} \| F_{nu} \, M_S^T \, \tilde{I} - b \|_2.
  \label{prob:mbrNonRect}
\end{equation}
Here, $\tilde{I}$ is a vector of those pixels within the FOV and $M_S$ is the linear transformation such that $M_S I$ is a vector of only those elements of $I$ that are within the FOV.  As before, if the data lies on a Cartesian grid, $F_{nu}$ can be replaced with $M_b\,F$.

Problem \eqref{prob:mbrNonRect} is a least-squares problem; it can either be solved analytically using the pseudo-inverse \cite{trefethen2022numerical}, or it can be solved numerically with LSQR \cite{paige1982lsqr}, which converges to a solution much faster than POCS.

For the case of parallel MRI, the image can be reconstructed by solving the following optimization problem:
\begin{equation}
    \text{minimize} \hspace{0.5em} \| \boldsymbol{F_{nu}} \, \boldsymbol{\sigma} \, M_S^T \, \tilde{I} - \boldsymbol{b} \|_2,
    \label{prob:mbrParallelMRI}
\end{equation}
where $\boldsymbol{\sigma}=\left[\sigma^{(1)}, \sigma^{(2)}, \ldots, \sigma^{(C)}\right]^T$, $\sigma^{(j)}$ is a diagonal matrix for the $j^{\text{th}}$ coil with diagonal elements equal to its sensitivity values, $C$ is the number of coils used for data collection, $\boldsymbol{F_{nu}}$ is a block-diagonal matrix where each block is $F_{nu}$, $\boldsymbol{b}=[b^{(1)}, b^{(2)}, \ldots, b^{(C)}]^T$, and $b^{(j)}$ is a vector of data collected by the $j^{\text{th}}$ coil.  Problem \eqref{prob:mbrParallelMRI} can also be solved with LSQR \cite{paige1982lsqr}.  Note that the additional information provided by the multiple coils can reduce the number of samples required for a high-quality image \cite{deshmane2012parallel}.

Note that problems \eqref{prob:mbrNonRect} and \eqref{prob:mbrParallelMRI} need not use the sampling pattern described in this manuscript.  These problems can be solved for any sampling pattern (e.g., spiral \cite{ahn1986high,meyer1992fast} or rosette \cite{noll1997multishot,bush2020rosette}).

\section{Experiments}
\label{sec:experiments}

Section \ref{sec:results} will present results of experiments involving MRI data of a sagittal slice of an ankle, an axial slice of a pineapple, and axial slices of a brain.  For all of the data presented, the data was collected with two dimensions of phase-encodes and one dimension of readout.  The data was then inverse Fourier transformed along the readout direction to place the data into the $k_x,k_y,z$ hybrid domain \cite{beatty2007method} (where $x,y,z$ are the spatial coordinates and $k_x,k_y,k_z$ are the frequency coordinates).  Once in the hybrid domain, the slice of each $z$ location is reconstructed independently.  Each dataset was normalized so that the maximum Fourier value had a magnitude of $1$.

MRI Data of the pineapple and brain were collected with a 3 Tesla clinical MRI machine for the purposes of the research presented in this manuscript.  All procedures performed in studies involving human participants were in accordance with the ethical standards of the institutional and/or national research committee and with the 1964 Helsinki declaration and its later amendments or comparable ethical standards.  Data were collected with health insurance portability and accountability act (HIPPA) compliance and we obtained prospective informed consent from the volunteer under COMIRB \#19-0158.

\begin{figure}[ht]
  \centering{}
  \includegraphics[width=0.95\linewidth]{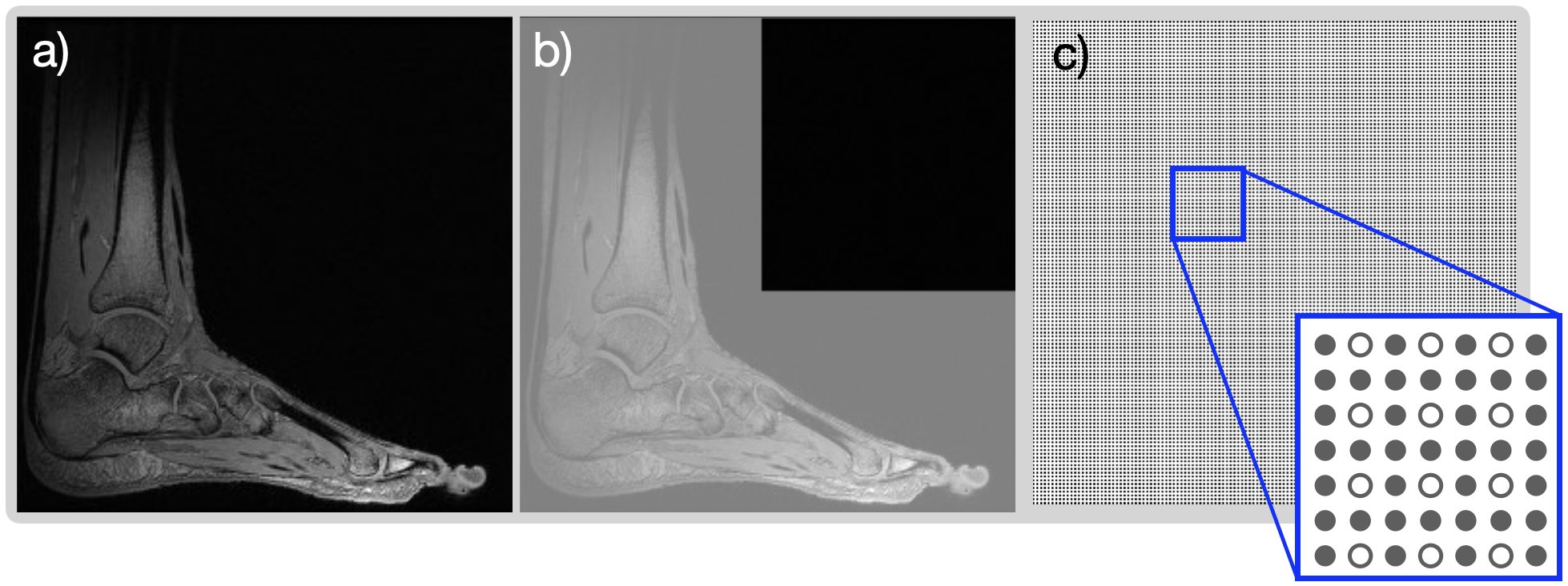}
  \caption{ Subfigure (a) shows a sagittal slice of an ankle and foot; (b) shows a corresponding non-rectangular FOV that does not include the upper-right quadrant, and (c) shows the corresponding sampling pattern for the non-rectangular FOV of (b).  The sampling burden of the reduced sampling pattern is $75\%$. }
  \label{fig:ankleFOV}
\end{figure}

The data of the ankle was collected with a 3D Cartesian full sampling pattern with two dimensions of phase-encodes and one dimension of readout (Fig. \ref{fig:ankleFOV}).  The data was collected using a 3 Tesla clinical MRI machine with a dedicated ankle coil array and was shared publicly as part of \cite{dwork2023accelerated}.  For the ankle, we will first simulate a single-coil acquisition: we will reconstruct the image from the fully-sampled data for all coils and we will combine those images using the method of \cite{roemer1990nmr}.  This will be considered the \textit{true} image.  We will Fourier transform this image to generate the data of the full sampling pattern, and we will eliminate 25\% of the data according to the sampling pattern of Fig. \ref{fig:ankleFOV}c for reconstruction with the non-rectangular FOV of Fig. \ref{fig:ankleFOV}b.  Separately, we will perform a parallel MRI reconstruction where we will simulate the the data that would have been acquired with the reduced sampling pattern.  To do so, we will first reconstruct the image of each coil from the fully-sampled data; we will then use inverse gridding to estimate the Fourier values for the spatial frequencies of the reduced sampling pattern.

The data of the pineapple and the brain were collected using a 3 Tesla clinical MRI machine with a birdcage coil array.  For these datasets, both a fully-sampled 3D Cartesian dataset and a prospectively undersampled dataset with the sampling pattern described for a non-rectangular FOV were collected.

Initially, for the ankle, we will use a non-rectangular FOV that consists of the 2nd, 3rd, and 4th quadrants.  That is, the FOV will not include the 1st quadrant, as depicted in Fig. \ref{fig:ankleFOV}.  For this FOV, the sampling pattern consists of every other row and every other column of the fully-sampled pattern; that is, in every $2\times 2$ block of samples three samples are included in the sampling pattern, as shown in Fig. \ref{fig:ankleFOV}c.  This will allow us to compare the results with \textit{retrospective downsampling}.  That is, we will reconstruct the image from the fully-sampled pattern and then reconstruct the image using only the samples from the sampling pattern of Fig. \ref{fig:ankleFOV}c.

\section{Results}
\label{sec:results}

For the following results, let the \textit{sampling burden} be the ratio of the number of samples of the given pattern to the number of samples of the full sampling pattern.  When coil sensitivities were required, they were estimated with fully-sampled data using the method of Pruessman et al. \cite{pruessmann1999sense}.  All data is normalized so that the Fourier value at the $0$ spatial frequency is $1$.

\begin{figure*}
  \centering{}
  \includegraphics[width=0.8\linewidth]{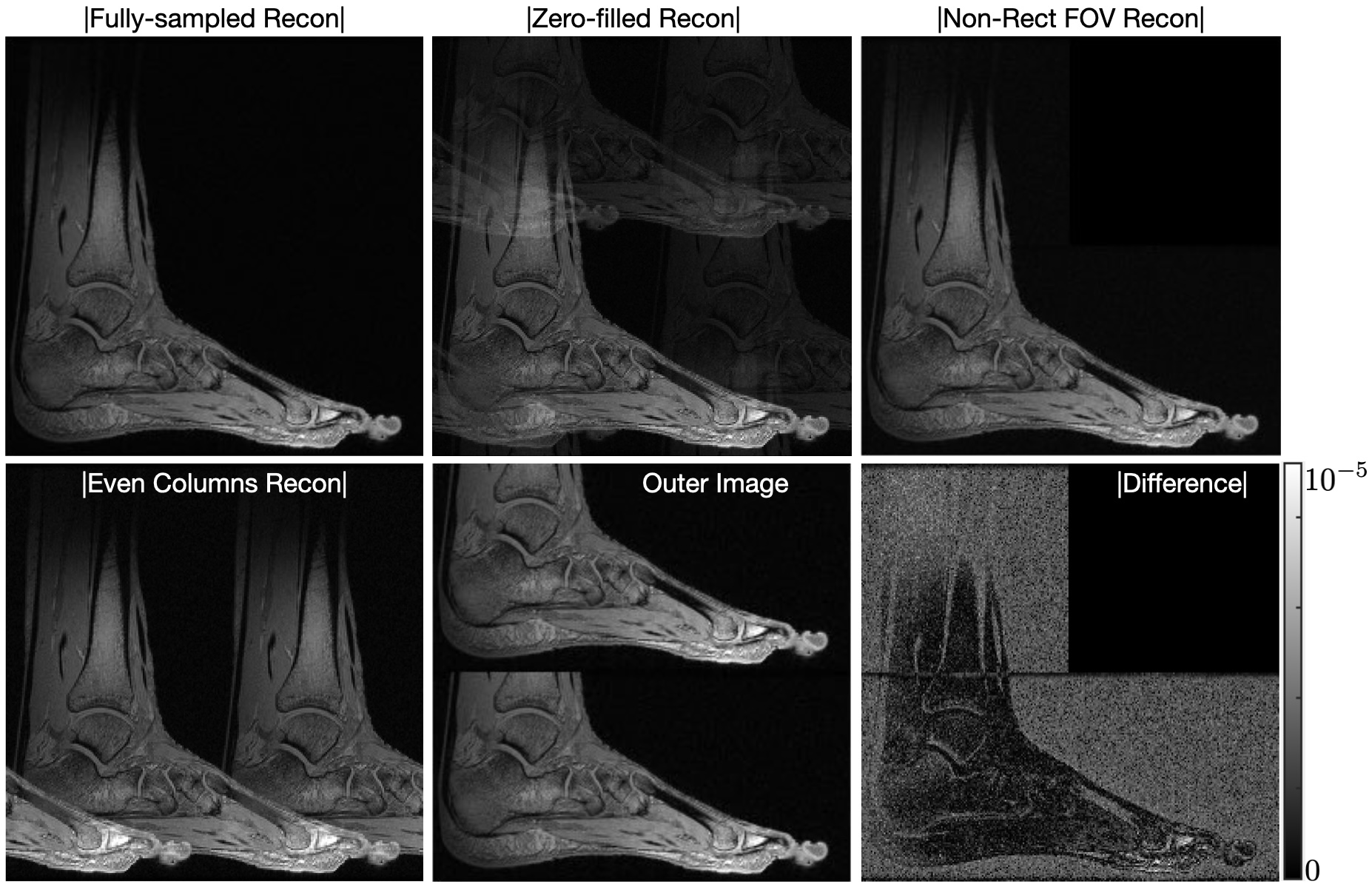}
  \caption{ Magnitude image reconstructions of a sagittal slice of an ankle and foot: a) shows the fully-sampled reconstruction, b) shows the zero-filled reconstruction with the non-rectangular FOV sampling pattern of Fig. \ref{fig:ankleFOV}b with a sampling burden of $75\%$, c) shows the reconstruction using Alg. \ref{alg:nonRectRecon}, c) the reconstruction using only the even columns of data, d) the reconstruction after subtracting away the Fourier values of the inner region, and e) the difference between (a) and (c). }
  \label{fig:ankleResultDirect}
\end{figure*}

Figure \ref{fig:ankleResultDirect} shows the direct reconstruction for the ankle from the simulated single-coil acquisition.  The magnitude of the difference ranges up to $10^{-5}$.  The sampling pattern for the non-rectangular FOV has a burden of $75\%$.

\begin{figure*}
  \centering{}
  \includegraphics[width=0.8\linewidth]{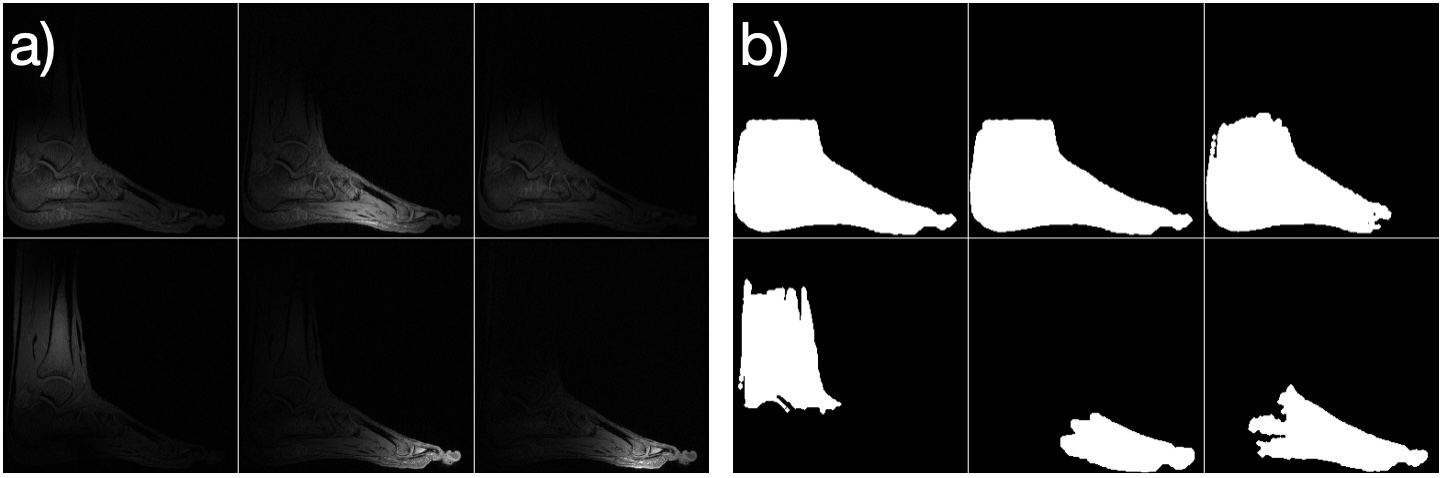}
  \caption{ Magnitude image reconstructions for six coils of a dedicated ankle array - a) Magnitude images for each coil, and b) the non-rectangular FOV determined using the intensity images of (a). }
  \label{fig:ankleCoilImgs}
\end{figure*}

Figure \ref{fig:ankleCoilImgs} shows the individual coil images from six coils of the ankle array and the corresponding supports of each image determined based on signal intensity.  The sampling pattern determined using these supports, as described in Sec. \ref{sec:methods}, has a sampling burden of $63\%$.

\begin{figure}[ht]
  \centering{}
  \includegraphics[width=0.95\linewidth]{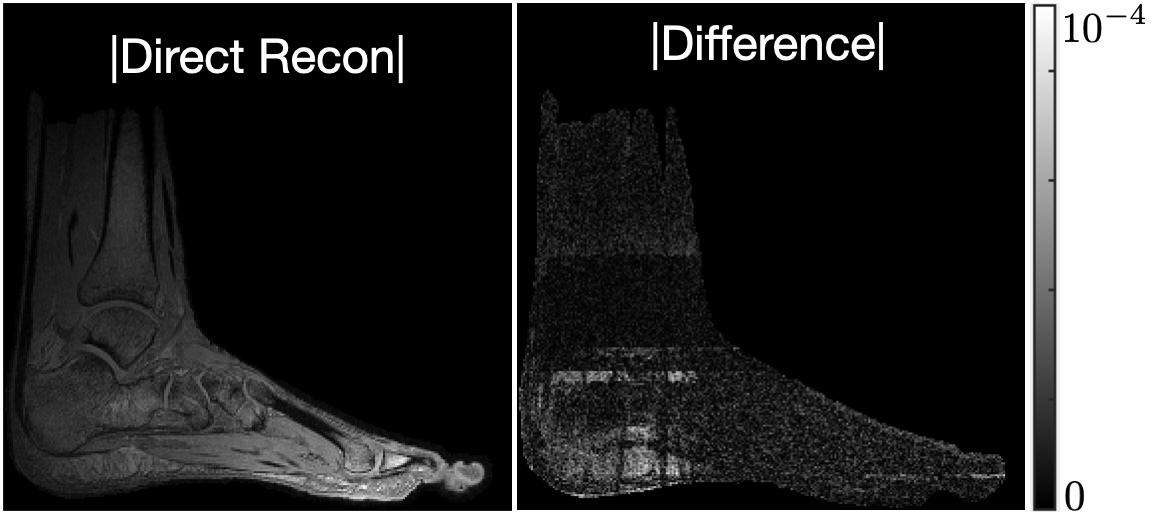}
  \caption{ Magnitude image reconstruction of the ankle using the direct reconstruction methods and the sampling pattern determined according to the individual coil FOVs shown in Fig. \ref{fig:ankleCoilImgs}b with a sampling burden of $63\%$.  The difference image shows the type of the errors that arise from this method; these errors are due to the fact that individual coils sense the whole subject imaged. }
  \label{fig:directAnkleParallelMRI}
\end{figure}

Figure \ref{fig:directAnkleParallelMRI} shows the reconstruction along with its difference from the fully-sampled reconstruction.  This reconstruction was accomplished with a sampling burden of $53\%$.  The difference with the fully-sampled reconstruction ranges up to $10^{-5}$.  The larger errors are due to the assumption of the smaller supports.  In truth, each coil senses the entire ankle, which leads to the errors in the results.

\begin{figure*}
  \centering{}
  \includegraphics[width=0.8\linewidth]{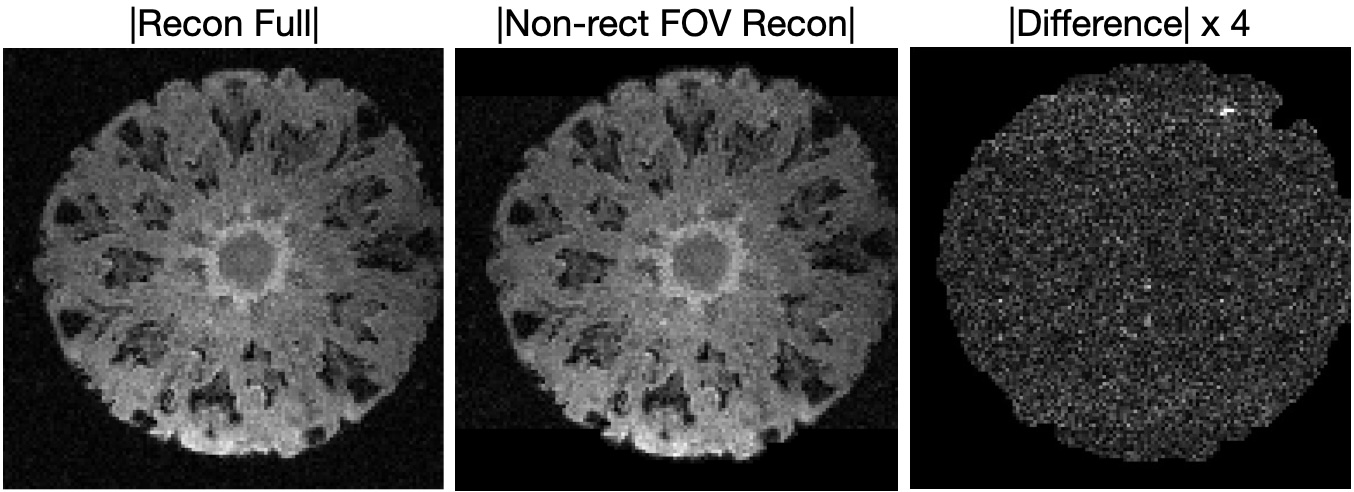}
  \caption{ Magnitude image reconstructions of an axial slice of a pineapple from data collected with a single coil: left) the fully sampled reconstruction, center) the direct reconstruction from the undersampled pattern with a burden of $87.5\%$, and right) the scaled magnitude of the difference. }
  \label{fig:pineappleDirect}
\end{figure*}

Figure \ref{fig:pineappleDirect} shows the direct reconstruction with a single coil for an axial slice of a pineapple.  The non-rectangular FOV was accomplished with an $87.5\%$ sampling burden.

\begin{figure*}[ht]
  \centering{}
  \includegraphics[width=0.85\linewidth]{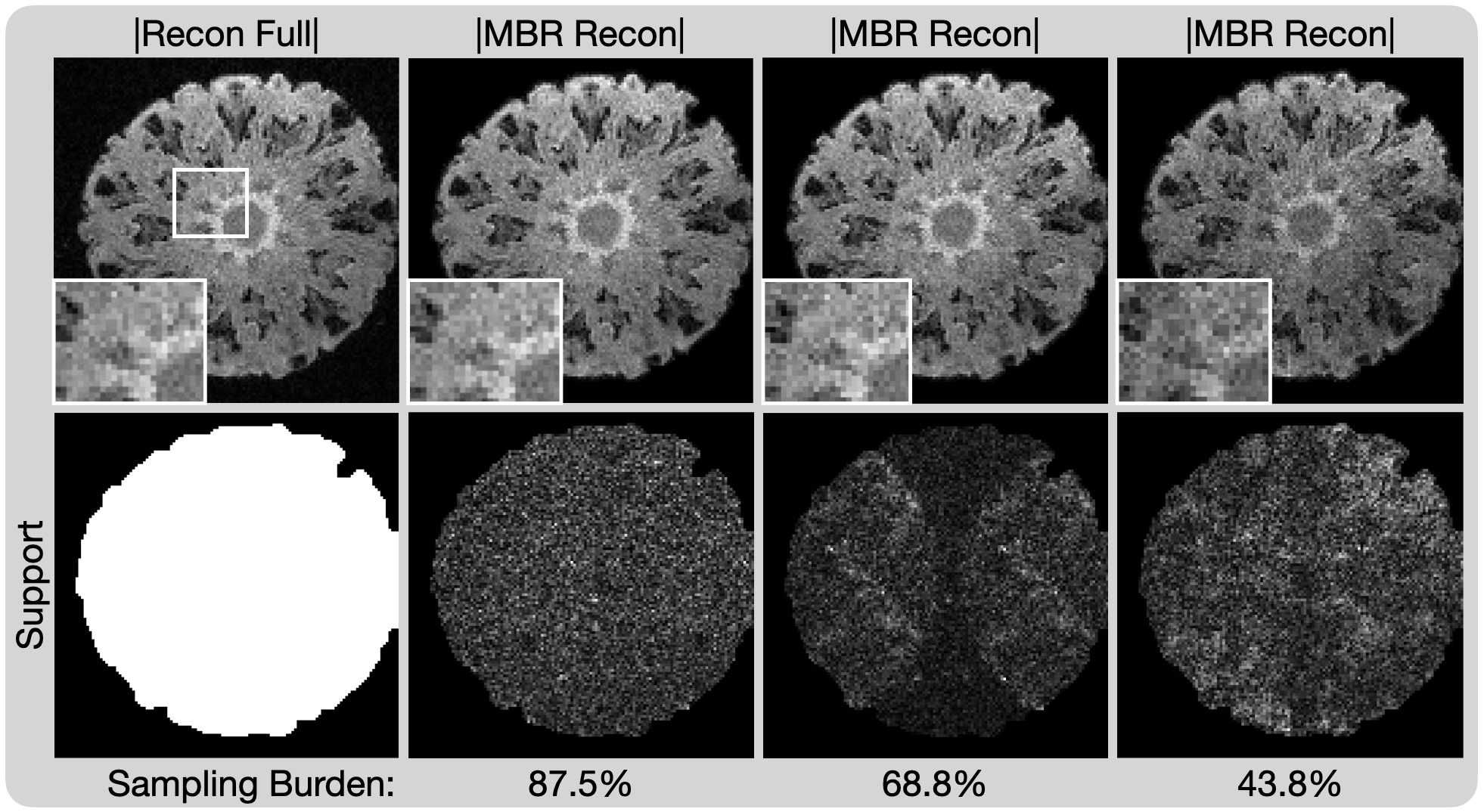}
  \caption{ Model-based magnitude image reconstructions of an axial slice of a pineapple with sampling burdens of (from left to right) $100\%$, $87.5\%$, $68.8\%$, and $43.8\%$. }
  \label{fig:pineappleMBR}
\end{figure*}

Figure \ref{fig:pineappleMBR} shows the model-based reconstruction of the axial slice of a pineapple with data collected from a 14 coil array.  Results are presented with sampling burdens of $87.5\%$, $68.8\%$, and $43.8\%$.  The sampling pattern was reduced by eliminating alternating samples from the even columns, and then eliminating alternating samples from the odd columns.  The mean squared error of the undersampled reconstructions are $1.6\times 10^{-9}$, $3.0\times 10^{-9}$, and $7.1\times 10^{-9}$, respectively.

\begin{figure*}[ht]
  \centering{}
  \includegraphics[width=0.85\linewidth]{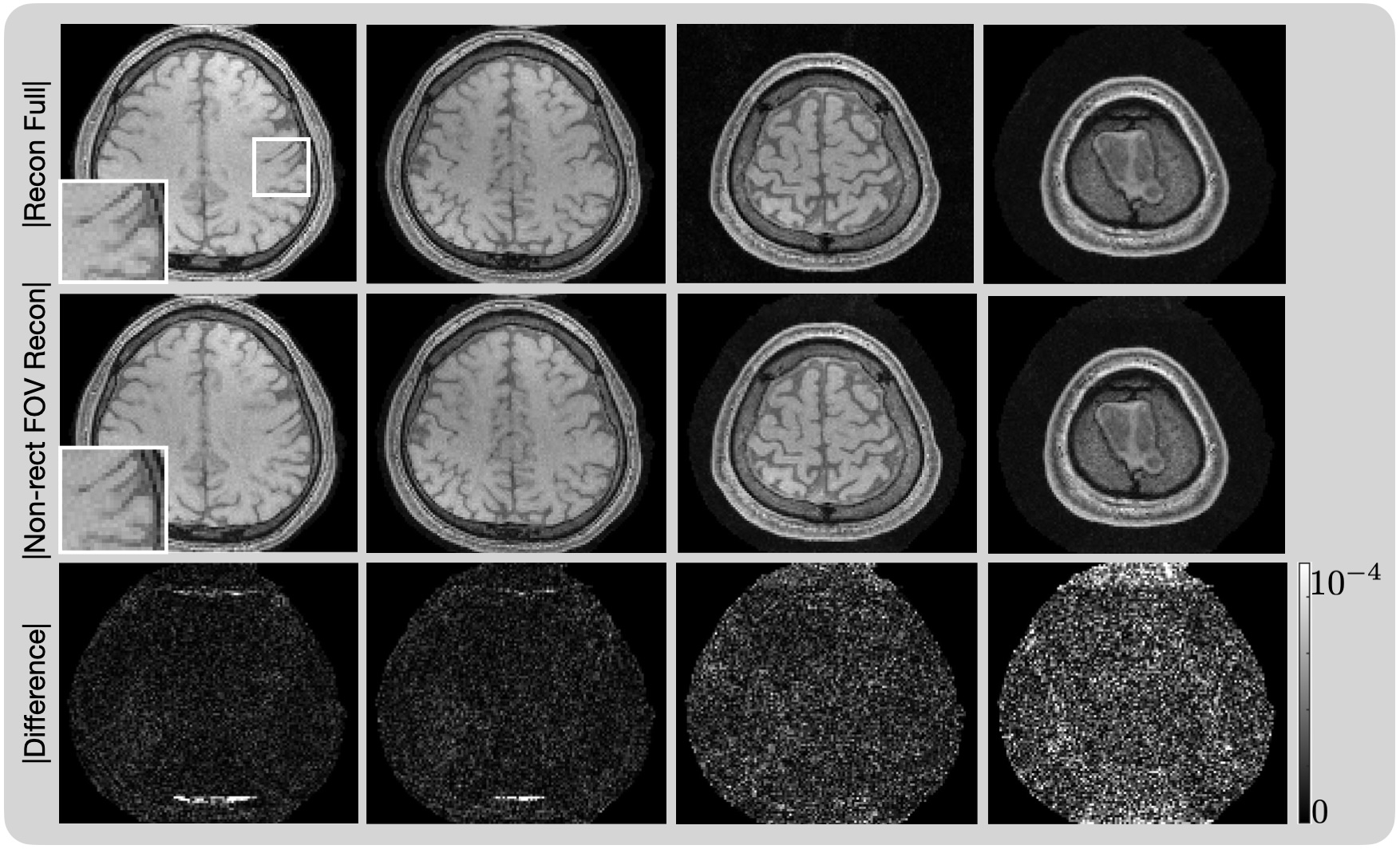}
  \caption{ Magnitude image reconstructions using the direct method for four axial slices of a brain using data collected with an $8$ coil array.  The sampling pattern of the non-rectangular FOV had a burden of 95\%. }
  \label{fig:brainSlices}
\end{figure*}

Figure \ref{fig:brainSlices} shows reconstructions using the non-rectangular FOV and compares them to the fully-sampled reconstructions for several slices of a human brain.  The sampling pattern of the non-rectangular FOV had a $95\%$ sampling burden.  The differences for all slices range from $0$ to $10^{-4}$.  As in all other examples, taking advantage of the non-rectangular FOV reconstructs a high-quality image with a reduced sampling burden.  The significant errors in the reconstruction are due to subject motion in between the full sampling pattern and the reduced sampling pattern; this would largely be eliminated if data were collected during a single scanning protocol.

\section{Conclusion}
\label{sec:conclusion}

The method presented in this manuscript takes advantage of a non-rectangular FOV with Fourier sensing to reconstruct a high-quality image with fewer samples.  We presented both a direct method of reconstruction as well as an iterative method for reconstruction.  Note that Gridding and inverse Gridding both take advantage of the Fast Fourier transform and are $\mathcal{O}(n\log n)$ algorithms (where $n$ is the number of elements of the input).  Since the direct reconstruction algorithm takes advantage of Gridding and Inverse Gridding, it too has a computational complexity of $\mathcal{O}(n\log n)$. 

The presented method reconstructs the image accurately as long as the non-rectangular FOV is a superset of the object's support.  Pixels within the FOV may be absent of any object.  However, to avoid errors, all pixels exterior to the FOV must be absent of any object.

In this manuscript, we also presented an iterative method of reconstructing an image with a non-rectangular FOV.  By incorporating this FOV into the objective function, we were able to eliminate the need for a projection operator and use a more efficient numerical solver than previous methods.  We further showed that we could reconstruct a high-quality image with even fewer samples in the parallel MRI setting.

Regularization terms can be added into the objective functions of \eqref{prob:mbrNonRect} and \eqref{prob:mbrParallelMRI}, which could permit high-quality reconstructions with even fewer samples.  Thus, one could incorporate compressed sensing \cite{candes2008introduction, lustig2007sparse}, perhaps with structured sparsity \cite{dwork2021utilizing, dwork2022utilizing}, into the reconstruction.

Note that once the Fourier values of the inner region are estimated at the spatial frequencies in $k_{\text{inner}}$, the data lies on a Caretsian grid.  This indicates that, rather than using gridding to reconstruct $I_{\text{inner}}$, one could instead use the inverse DFT, which would be more efficient.  Note, though, that one would first have to apply a linear phase ramp to the Fourier values so that the reconstructed image is shifted appropriately to the rows of the inner region.

In this work, we collected the data for the sampling pattern of the non-rectangular FOV by combining data from two different data collections: a fully-sampled pattern and an under-sampled pattern with aliasing in the vertical dimension.  In the near future, we will implement a custom protocol onto the clinical scanner so that only the data required for the non-rectangular FOV is collected.

In this work, we assumed that the even columns would be the alternating columns of a full sampling pattern.  This led to aliasing by half the FOV as depicted in Fig. \ref{fig:imgParts}a.  Alternatively, one could permit an arbitrary amount of aliasing, which would correspond to different separations between the even columns.  By searching over the amount of aliasing, it is likely that one would find a sampling pattern with fewer samples than that presented in this manuscript.  In the same vein, one could try to identify a rotation of the sampling pattern that leads to a reduced sampling burden.  We leave this pursuit as future work.

MATLAB code for reconstructing images in accordance to the methods of this manuscript will be made available at \url{https://github.com/ndwork/nonRectSupport} and at the following website \url{www.nicholasdwork.com}.  The pineapple and brain data collected for this manuscript will also be made publicly available at \url{www.nicholasdwork.com}.




\begin{thebibliography}{10}

\bibitem{nishimura1996principles}
Dwight~G Nishimura.
\newblock {\em Principles of magnetic resonance imaging}.
\newblock lulu.com, 1996.

\bibitem{sumpf2011model}
Tilman~J Sumpf, Martin Uecker, Susann Boretius, and Jens Frahm.
\newblock Model-based nonlinear inverse reconstruction for {T2} mapping using
  highly undersampled spin-echo {MRI}.
\newblock {\em Journal of Magnetic Resonance Imaging}, 34(2):420--428, 2011.

\bibitem{kajbaf2013compressed}
Hamed Kajbaf, Joseph~T Case, Zengli Yang, and Yahong~Rosa Zheng.
\newblock Compressed sensing for sar-based wideband three-dimensional microwave
  imaging system using non-uniform fast fourier transform.
\newblock {\em IET Radar, Sonar \& Navigation}, 7(6):658--670, 2013.

\bibitem{baron2018rapid}
Corey~A Baron, Nicholas Dwork, John~M Pauly, and Dwight~G Nishimura.
\newblock Rapid compressed sensing reconstruction of {3D} non-cartesian {MRI}.
\newblock {\em Magnetic resonance in medicine}, 79(5):2685--2692, 2018.

\bibitem{cole2021fast}
Elizabeth~K Cole, Frank Ong, Shreyas~S Vasanawala, and John~M Pauly.
\newblock Fast unsupervised {MRI} reconstruction without fully-sampled ground
  truth data using generative adversarial networks.
\newblock In {\em Proceedings of the IEEE/CVF International Conference on
  Computer Vision}, pages 3988--3997, 2021.

\bibitem{dudgeon1983multidimensional}
Dan~E Dudgeon.
\newblock Multidimensional digital signal processing.
\newblock {\em Engewood Cliffs}, 1983.

\bibitem{bracewell1995two}
Ronald~N Bracewell.
\newblock {\em Two-dimensional imaging}.
\newblock Prentice-Hall, Inc., 1995.

\bibitem{birdsong2016hexagonal}
James~B Birdsong and Nicholas~I Rummelt.
\newblock The hexagonal fast fourier transform.
\newblock In {\em IEEE International Conference on Image Processing (ICIP)},
  pages 1809--1812. IEEE, 2016.

\bibitem{samsonov2004pocsense}
Alexei~A Samsonov, Eugene~G Kholmovski, Dennis~L Parker, and Chris~R Johnson.
\newblock {POCSENSE}: {POCS}-based reconstruction for sensitivity encoded
  magnetic resonance imaging.
\newblock {\em Magnetic Resonance in Medicine: An Official Journal of the
  International Society for Magnetic Resonance in Medicine}, 52(6):1397--1406,
  2004.

\bibitem{fessler2010model}
Jeffrey~A Fessler.
\newblock Model-based image reconstruction for {MRI}.
\newblock {\em IEEE Signal Processing Magazine}, 27(4):81--89, 2010.

\bibitem{greengard2004accelerating}
Leslie Greengard and June-Yub Lee.
\newblock Accelerating the nonuniform fast fourier transform.
\newblock {\em SIAM review}, 46(3):443--454, 2004.

\bibitem{jackson1991selection}
John~I Jackson, Craig~H Meyer, Dwight~G Nishimura, and Albert Macovski.
\newblock Selection of a convolution function for fourier inversion using
  gridding (computerised tomography application).
\newblock {\em IEEE transactions on medical imaging}, 10(3):473--478, 1991.

\bibitem{beatty2005rapid}
Philip~J Beatty, Dwight~G Nishimura, and John~M Pauly.
\newblock Rapid gridding reconstruction with a minimal oversampling ratio.
\newblock {\em IEEE transactions on medical imaging}, 24(6):799--808, 2005.

\bibitem{dwork2023optimization}
Nicholas Dwork, Daniel O'Connor, Ethan~MI Johnson, Corey~A Baron, Jeremy~W
  Gordon, John~M Pauly, and Peder~EZ Larson.
\newblock Optimization in the space domain for density compensation with the
  nonuniform {FFT}.
\newblock {\em Magnetic Resonance Imaging}, 100:102--111, 2023.

\bibitem{pauly2005nonCartesian}
John Pauly.
\newblock Non-{Cartesian} reconstruction.
\newblock {\em preprint}, 2005.

\bibitem{rasche1999resampling}
Volker Rasche, Roland Proksa, R~Sinkus, Peter Bornert, and Holger Eggers.
\newblock Resampling of data between arbitrary grids using convolution
  interpolation.
\newblock {\em IEEE transactions on medical imaging}, 18(5):385--392, 1999.

\bibitem{roemer1990nmr}
Peter~B Roemer, William~A Edelstein, Cecil~E Hayes, Steven~P Souza, and
  Otward~M Mueller.
\newblock The {NMR} phased array.
\newblock {\em Magnetic resonance in medicine}, 16(2):192--225, 1990.

\bibitem{trefethen2022numerical}
Lloyd~N Trefethen and David Bau.
\newblock {\em Numerical linear algebra}.
\newblock SIAM, 1997.

\bibitem{paige1982lsqr}
Christopher~C Paige and Michael~A Saunders.
\newblock {LSQR}: An algorithm for sparse linear equations and sparse least
  squares.
\newblock {\em ACM Transactions on Mathematical Software (TOMS)}, 8(1):43--71,
  1982.

\bibitem{deshmane2012parallel}
Anagha Deshmane, Vikas Gulani, Mark~A Griswold, and Nicole Seiberlich.
\newblock Parallel {MR} imaging.
\newblock {\em Journal of Magnetic Resonance Imaging}, 36(1):55--72, 2012.

\bibitem{ahn1986high}
CB~Ahn, JH~Kim, and ZH~Cho.
\newblock High-speed spiral-scan echo planar {NMR} imaging-{I}.
\newblock {\em IEEE transactions on medical imaging}, 5(1):2--7, 1986.

\bibitem{meyer1992fast}
Craig~H Meyer, Bob~S Hu, Dwight~G Nishimura, and Albert Macovski.
\newblock Fast spiral coronary artery imaging.
\newblock {\em Magnetic resonance in medicine}, 28(2):202--213, 1992.

\bibitem{noll1997multishot}
Douglas~C Noll.
\newblock Multishot rosette trajectories for spectrally selective {MR} imaging.
\newblock {\em IEEE transactions on medical imaging}, 16(4):372--377, 1997.

\bibitem{bush2020rosette}
Adam~M Bush, Christopher~M Sandino, Shreya Ramachandran, Frank Ong, Nicholas
  Dwork, Evan~J Zucker, Ali~B Syed, John~M Pauly, Marcus~T Alley, and Shreyas~S
  Vasanawala.
\newblock Rosette trajectories enable ungated, motion-robust, simultaneous
  cardiac and liver t2* iron assessment.
\newblock {\em Journal of Magnetic Resonance Imaging}, 52(6):1688--1698, 2020.

\bibitem{beatty2007method}
PJ~Beatty, AC~Brau, S~Chang, S~Joshi, CR~Michelich, E~Bayram, TE~Nelson,
  RJ~Herfkens, and JH~Brittain.
\newblock A method for autocalibrating {2D} accelerated volumetric parallel
  imaging with clinically practical reconstruction times.
\newblock In {\em Proceedings of the International Society for Magnetic
  Resonance in Medicine}, volume~15, page 1749, 2007.

\bibitem{dwork2023accelerated}
Nicholas Dwork and Erin~K Englund.
\newblock Accelerated parallel magnetic resonance imaging with compressed
  sensing using structured sparsity.
\newblock {\em arXiv preprint arXiv:2312.01610}, 2023.

\bibitem{pruessmann1999sense}
Klaas~P Pruessmann, Markus Weiger, Markus~B Scheidegger, and Peter Boesiger.
\newblock {SENSE}: sensitivity encoding for fast {MRI}.
\newblock {\em Magnetic Resonance in Medicine: An Official Journal of the
  International Society for Magnetic Resonance in Medicine}, 42(5):952--962,
  1999.

\bibitem{candes2008introduction}
Emmanuel~J Cand{\`e}s and Michael~B Wakin.
\newblock An introduction to compressive sampling.
\newblock {\em IEEE signal processing magazine}, 25(2):21--30, 2008.

\bibitem{lustig2007sparse}
Michael Lustig, David Donoho, and John~M Pauly.
\newblock Sparse {MRI}: The application of compressed sensing for rapid {MR}
  imaging.
\newblock {\em Magnetic Resonance in Medicine: An Official Journal of the
  International Society for Magnetic Resonance in Medicine}, 58(6):1182--1195,
  2007.

\bibitem{dwork2021utilizing}
Nicholas Dwork, Daniel O’Connor, Corey~A Baron, Ethan~MI Johnson, Adam~B
  Kerr, John~M Pauly, and Peder~EZ Larson.
\newblock Utilizing the wavelet transform’s structure in compressed sensing.
\newblock {\em Signal, image and video processing}, 15:1407--1414, 2021.

\bibitem{dwork2022utilizing}
Nicholas Dwork and Peder~EZ Larson.
\newblock Utilizing the structure of a redundant dictionary comprised of
  wavelets and curvelets with compressed sensing.
\newblock {\em Journal of Electronic Imaging}, 31(6):063043--063043, 2022.

\end{thebibliography}

\end{document}